# THE SINGLE-PARTICLE STRUCTURE OF NEUTRON-RICH NUCLEI OF ASTROPHYSICAL INTEREST AT THE ORNL HRIBF[*]

D.W. BARDAYAN[1], J.C. BATCHELDER[2], J.C. BLACKMON[1], C.R. BRUNE[3], A.E. CHAMPAGNE[4], J.A. CIZEWSKI[5], T. DAVINSON[6], U. GREIFE[7], A.N. JAMES[8], M. JOHNSON[5], R.L. KOZUB[9], J.F. LIANG[1], R.J. LIVESAY[7], Z. MA[10], C.D. NESARAJA[9], D.C. RADFORD[1], D. SHAPIRA[1], M.S. SMITH[1], J. S. THOMAS[5], P.J. WOODS[6], E. ZGANJAR[11], AND THE UNIRIB COLLABORATION

[1]*Physics Division, Oak Ridge National Laboratory[*], Oak Ridge, TN 37831 USA*
[2]*UNIRIB[†], Oak Ridge Associated Universities, Oak Ridge, TN 37831 USA*
[3]*Dept. of Physics and Astronomy, Ohio University, Athens, OH 45701 USA*
[4]*Dept. of Physics and Astronomy, Univ. of North Carolina, Chapel Hill, NC 27599 USA*
[5]*Dept. of Physics and Astronomy, Rutgers University, New Brunswick, NJ 08903 USA*
[6]*Dept. of Physics, Univ. of Edinburgh, Edinburgh EH9 3JZ United Kingdom*
[7]*Dept. of Physics, Colorado School of Mines, Golden, CO 80401 USA*
[8]*Physics Dept., University of Liverpool, Liverpool L693BX United Kingdom*
[9]*Dept. of Physics, Tennessee Technological University, Cookeville, TN 38505 USA*
[10]*Dept. of Physics and Astronomy, University of Tennessee, Knoxville, TN 37996 USA*
[11]*Dept. of Physics and Astronomy, Louisiana State Univ., Baton Rouge, LA 70803 USA*

The rapid neutron-capture process ($r$ process) produces roughly half of the elements heavier than iron. The path and abundances produced are uncertain, however, because of the lack of nuclear structure information on important neutron-rich nuclei. We are studying nuclei on or near the r-process path via single-nucleon transfer reactions on neutron-rich radioactive beams at ORNL's Holifield Radioactive Ion Beam Facility (HRIBF). Owing to the difficulties in studying these reactions in inverse kinematics, a variety of experimental approaches are being developed. We present the experimental methods and initial results.

## 1. Introduction

Almost all of the elements heavier than lithium populating the Cosmos were created in stellar burning or explosions [1]. One important source is the astrophysical $r$ process which is thought to have created roughly half of the elements heavier than iron [2]. Heavy nuclei are produced in the $r$ process by a series of neutron captures and $\beta$ decays that flows through extremely neutron-rich nuclei ending near uranium. Abundances recently measured in some very old stars are consistent with an *r*-process origin [3]. Comparisons of measured abundances with model calculations can constrain the astrophysical conditions in which the *r* process occurs (e.g., can it occur in supernovae) and even the age of the Galaxy. The *r*-process models are currently very uncertain, however, because the nuclear structure of most *r*-process nuclei is not known. Especially important is the structure of nuclei near closed neutron shells where the *r*-process abundances peak as a result of the small neutron-capture cross sections of these nuclei. A recent study [4] found that order of magnitude variations in the predicted *r*-process abundances result from uncertainties in the estimated neutron-induced reaction cross sections. Better nuclear structure information is needed on these neutron-rich nuclei to better constrain nuclear structure and mass models, improve cross section estimates, and in turn improve the *r*-process abundance calculations.

---

[*] ORNL is managed by UT-Battelle, LLC for the U.S. Dept. of Energy under contract DE-AC05-00OR22725.
[†] UNIRIB is a consortium of universities, the state of TN, Oak Ridge Associated Universities, and ORNL and is partially supported by them.

Accelerated beams of many neutron-rich nuclei have recently become available at ORNL's Holifield Radioactive Ion Beam Facility [5]. These nuclei are produced by proton-induced fission of uranium, which has been deposited as UC on a low-density graphite matrix. The radioactive nuclei are then accelerated to energies up to 4-5 MeV/amu by the 25-MV Tandem accelerator. The available neutron-rich beams with intensities greater than $10^3$ ions/second are shown as gray boxes in Fig. 1. Near the neutron closed shells, N = 50 and 82, available beams include nuclei on or near the r-process path. Single-nucleon transfer reactions are being used with these beams to study the structure of these neutron-rich nuclei of astrophysical importance.

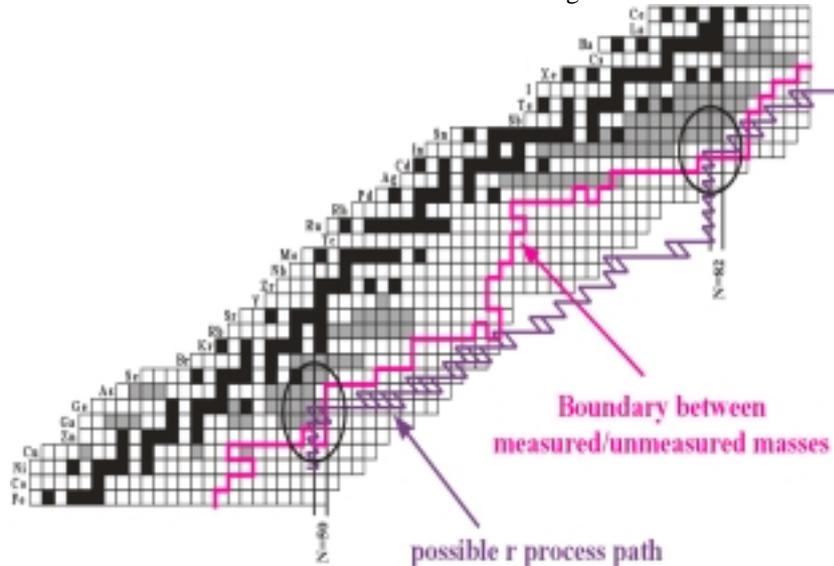

Figure 1. The available neutron-rich beams are shown as gray boxes. Stable isotopes are shown as black boxes. A possible r-process path is also shown. Near the neutron closed shells N=50 and 82, studies of r-process nuclei are possible.

## 2. Neutron-Transfer Reactions

Single-neutron transfer reactions are being studied via (d,p) or ($^9$Be,$^8$Be) reactions in inverse kinematics. The excitation energies of single-particle neutron states are determined either from the proton-energy spectrum at a fixed angle or from the energies of the γ-rays emitted when the excited nucleus decays. Information on spectroscopic factors and the spins of populated states are also obtained from the measured (d,p) cross sections and angular distributions. The kinematics of such reactions, however, present difficult challenges that require multiple experimental approaches to overcome. As an example, we plot in Fig. 2 the emitted proton energies calculated for d($^{132}$Sn,p)$^{133}$Sn at 5 MeV/amu as a function of their laboratory angles. The forward center-of-mass angles, at which the angular distributions are most sensitive to the transferred angular momentum, correspond to backward laboratory angles. At these angles, however, the proton energies for excited states rapidly drop below what is reasonable for particle identification with a silicon-detector telescope. The (d,p) cross section is also much smaller at backward laboratory angles than at forward angles. The best solution typically is to cover a relatively large angular range ($\theta_{lab}$=70°-130°) with two-dimensional position-sensitive silicon-strip detector telescopes. The two-dimensional position sensitivity is required to determine the reaction angle of the detected protons upon which the proton energy depends sensitively. A typical target-detector configuration is pictured in Fig. 3 [6]. In a recent stable beam test [7], beams of ~4 MeV/amu $^{124}$Sn bombarded 200 μg/cm$^2$ CD$_2$ targets rotated 60° to the beam axis. Protons from the d($^{124}$Sn,p)$^{125}$Sn [7] reaction were distinguished from scattered target ions by standard energy-loss techniques (Fig. 3). The identification of protons was sufficient to identify the events of interest as the yield of protons from other reactions with contaminants in the target was found to be negligible. Protons populating several states in $^{125}$Sn were identified from the plot of proton energy versus laboratory angle (Fig. 3).

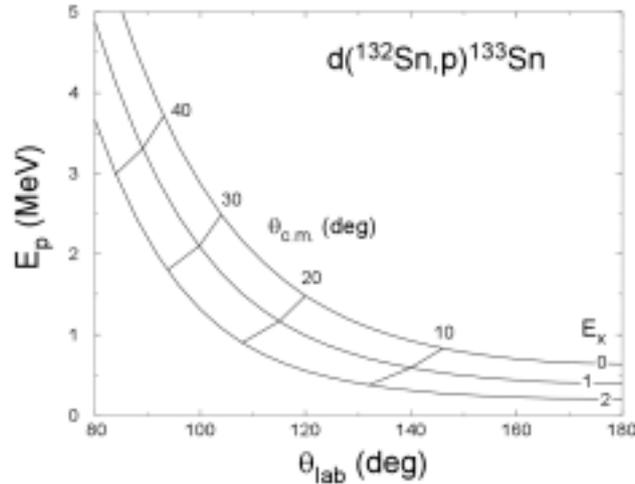

Figure 2. Reaction kinematics for d($^{132}$Sn,p)$^{133}$Sn in inverse kinematics at 5 MeV/amu.

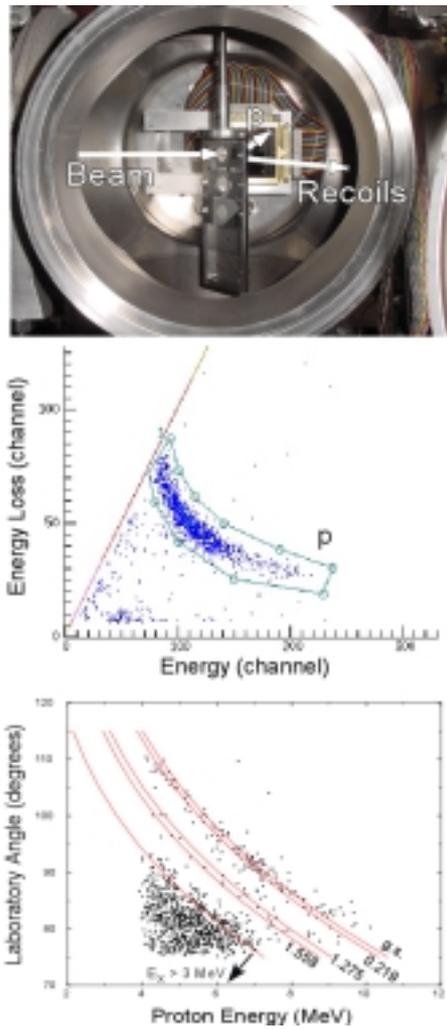

Figure 3. A typical detector setup is shown along with particle identification, and proton energies from the d($^{124}$Sn,p)$^{125}$Sn reaction [7] at 4 MeV/amu. Expected kinematical curves for protons populating $^{125}$Sn states (labeled in MeV) are shown. Excitation energies are from Ref. [8].

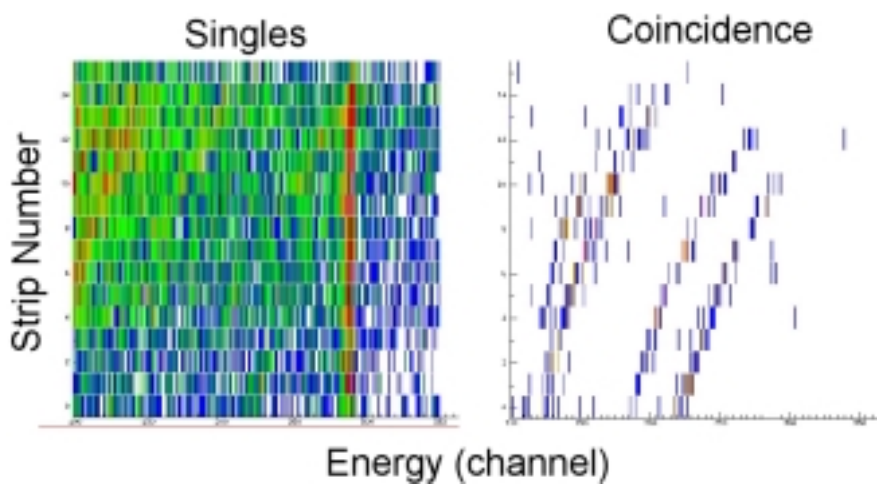

Figure 4. The energy spectrum of charged particles observed in a recent study of the d($^{18}$F,p)$^{19}$F reaction at 6 MeV/amu is plotted for each detector strip in singles and in coincidence with $^{19}$F ions at the focal plane of the DRS. Strips 0 - 15 correspond to laboratory angles from 157° - 118°, respectively.

In other cases to identify the events of interest, it will be necessary to detect the heavy recoil in coincidence with reaction protons. This can be accomplished by placing a microchannel plate [9] downstream of the target or by separating the recoils from beam particles using a mass separator such as the Daresbury Recoil Separator (DRS), which has been installed at the HRIBF for studies of reactions of astrophysical interest with radioactive ion beams [10]. This technique was used recently for a study of the d($^{18}$F,p)$^{19}$F reaction [11]. Protons from the d($^{18}$F,p)$^{19}$F reaction were detected at backward laboratory angles in coincidence with recoil $^{19}$F ions detected at the focal plane of the DRS. As shown in Fig. 4, protons populating states in $^{19}$F were readily identified when the coincidence requirement was applied.

Because of large kinematical shifts and high level densities, it may not be possible to achieve the necessary energy resolution in the charged-particle spectrum to resolve all the states in the compound nucleus. In these cases, the excitation energies of single-particle levels can be inferred from the energies of γ-rays observed following a neutron-transfer reaction. This method has been used to probe the level structure of $^{135}$Te using a radioactive $^{134}$Te beam and a $^9$Be target. The resulting $^8$Be ions from the $^9$Be($^{134}$Te,$^8$Be)$^{135}$Te reaction broke up into two α-particles that were detected in coincidence in single CsI crystals of the ORNL HYBALL detector [12]. As shown in Fig. 5, the detection of coincident α-particle pairs was an extremely clean signature of neutron transfer. The CLARION Ge detector array [12] was triggered on these 2α events finding several γ-ray transitions in $^{135}$Te.

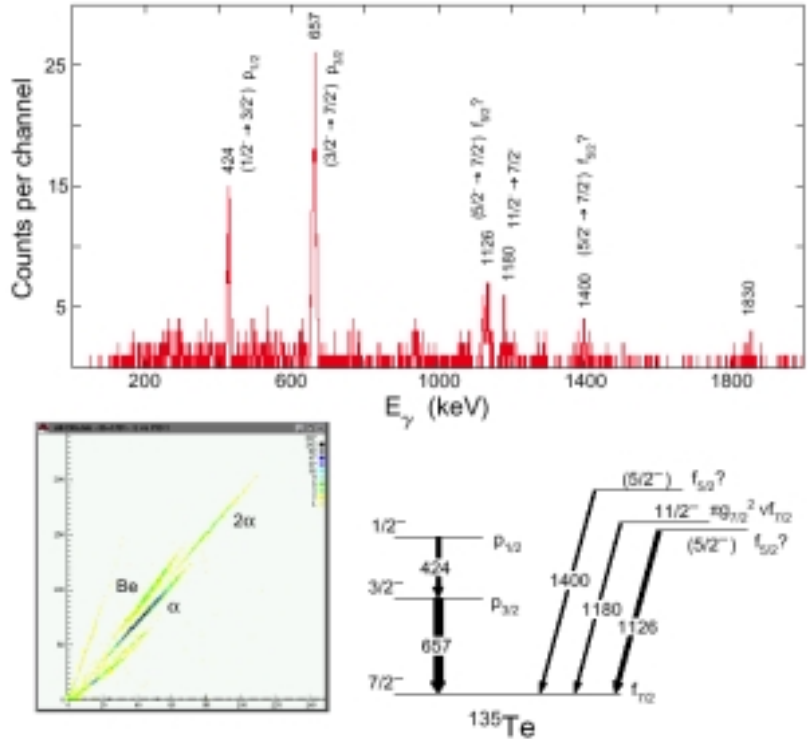

Figure 5. TOP: Gamma-ray spectrum from the $^9$Be($^{134}$Te,$^8$Be)$^{135}$Te reaction at 4 MeV/amu in coincidence with 2α clusters detected by the HYBALL. BOTTOM LEFT: Particle identification spectrum for one of the CsI detectors. BOTTOM RIGHT: Partial level scheme for $^{135}$Te, showing tentative configuration assignments.

### 3. Proton-Transfer Reactions

Proton-transfer studies on neutron-rich radioactive beams can also be used to probe the structure of nuclei near neutron closed shells. For instance, $^{131}$In is believed to be on the r-process path and is one proton away from doubly-magic $^{132}$Sn, an available radioactive beam at the HRIBF. For doubly-magic nuclei such as $^{132}$Sn, the high energy of core excitations limits the mixing of core-coupled states into the states of adjacent nuclides, and thus the single-particle levels of $^{131}$In should be relatively pure.

Knowledge of the energies of the single-particle levels in $^{131}$In would provide a much needed test of the shell model in this crucial mass region. The $\pi p_{3/2}$ and $\pi f_{5/2}$ hole states in $^{131}$In are important single-particle levels that are not observed in decay studies. They should, however, be observable using single-proton transfer reactions with a $^{132}$Sn beam. Similar to the method used for the $^{9}$Be($^{134}$Te,$^{8}$Be)$^{135}$Te study, 2$\alpha$ clusters would be detected in the HYBALL detector from the $^{7}$Li($^{132}$Sn,$^{8}$Be)$^{131}$In reaction [13]. The excitation energies of single-particle levels in $^{131}$In would be deduced from the $\gamma$-rays detected in the CLARION Ge array in coincidence with the 2$\alpha$ clusters.

### 4. Conclusions

The availability of neutron-rich radioactive beams at the HRIBF provides exciting opportunities for study of the structure of nuclei far from stability. Such information is critical to our understanding of the astrophysical *r* process. Especially important is the understanding of structure near neutron closed shells where the *r*-process abundances tend to peak. We are developing the techniques needed to study these important nuclei via single-nucleon transfer reactions using neutron-rich radioactive beams. The need to study the reactions in inverse kinematics presents many experimental difficulties that are being overcome by using a variety of techniques. The detection of protons from the (d,p) reaction in silicon-strip detectors in coincidence with heavy recoils provides a promising mechanism for the extraction of single-particle strengths and angular-momentum transfers. A complimentary technique is to use a single-particle transfer reaction resulting in the emission of a $^{8}$Be nucleus that would, in turn, break up into particle clusters providing a highly-selective trigger. The detection of these particle clusters in coincidence with $\gamma$-rays provides a good measure of the energies of single-particle levels in the resulting nucleus. In addition to providing important nuclear structure information, the techniques we are developing will also be useful when the Rare Isotope Accelerator (RIA) [14] comes online, and the study of reactions in inverse-kinematics on nuclei further from stability becomes possible.

**Acknowledgments**

We thank the staff of the HRIBF whose hard work made these experiments possible.